\documentclass[aps,prd,superscriptaddress,preprint
 reprint,
 amssymb,
 aps,
]{revtex4-2}
\newcommand{\nn}{\nonumber}
\usepackage{amsmath}
\usepackage{hyperref}
\usepackage{graphicx}
\usepackage{dcolumn}
\usepackage{bm}
\usepackage{array}
\usepackage{booktabs}
\usepackage{mathrsfs}
\usepackage{color}
\usepackage{soul}
\usepackage[normalem]{ulem}

\begin{document}

\preprint{APS/123-QED}

\title{Regularization Prescription for the Mixing Between Nonlocal Gluon and Quark Operators}

\author{Yao Ji}
\affiliation{School of Science and Engineering, The Chinese University of Hong Kong, Shenzhen 518172, China}

\author{Zhuoyi Pang}
 \email{corresponding author: Zhuoyi.Pang@ncbj.gov.pl}
\affiliation{School of Science and Engineering, The Chinese University of Hong Kong, Shenzhen 518172, China}
\affiliation{National Centre for Nuclear Research (NCBJ), 02-093 Warsaw, Poland}


\author{Fei Yao}
\affiliation{Physics Department, Brookhaven National Laboratory, Upton, New York 11973, USA
}
\affiliation{School of Science and Engineering, The Chinese University of Hong Kong, Shenzhen 518172, China}

\author{Jian-Hui Zhang}
\email{corresponding author: zhangjianhui@cuhk.edu.cn}
\affiliation{School of Science and Engineering, The Chinese University of Hong Kong, Shenzhen 518172, China}



\begin{abstract}
It is well-known that in the study of mixing between nonlocal gluon and quark
bilinear operators there exists an ambiguity when relating coordinate space and momentum
space results. In this work, we show that this ambiguity is due to the lack of a proper
regularization prescription of the singularity that arises when the separation between the
gluon/quark fields approaches zero. We then demonstrate that dimensional regularization
resolves this issue and yields consistent results in both coordinate and momentum space.
This prescription is also compatible with lattice extractions of parton distributions from
nonlocal operators.
\end{abstract}

\maketitle


\section{Introduction}
Parton physics provides a useful framework for studying the inner structure of hadrons through their fundamental constituents --- quarks and gluons. It is naturally formulated in light-front (LF) quantization, where nonlocal quark and gluon operators along lightlike separations (LF operators) play a crucial role. In the collinear case, these operators define fundamental parton quantities such as parton distribution functions (PDFs), generalized parton distributions (GPDs) and distribution amplitudes (DAs). A notable feature of collinear parton distributions is the mixing between flavor-singlet quark operators and gluon operators under QCD evolution. As the quark and gluon LF operators have different mass dimensions, the mixing coefficient has to incorporate the lightlike separation of the nonlocal operators to compensate for the dimensional mismatch. This mixing pattern has been extensively studied in both coordinate and momentum space (see, e.g.,~\cite{Ji:1996nm} for calculation in momentum space,~\cite{Blumlein:1997pi,Balitsky:1987bk} for calculation in coordinate space, and~\cite{Belitsky:2005qn,Diehl:2003ny} for reviews), revealing important insights into the scale dependence of collinear parton distributions. However, there exists a long-standing ambiguity when relating coordinate space and momentum space results in the gluon-in-quark channel. Since the gluon LF operator has mass dimension 4 whereas the quark LF operator has mass dimension 3, the mixing coefficient is inversely proportional to the lightcone distance. When Fourier transforming the coordinate space result to momentum space, a conventional approach is to replace the former with an integral form, but with an indefinite integration limit~\cite{Radyushkin:1997ki,Belitsky:2005qn,Yao:2022vtp}. This leads to ambiguities in the Fourier transformed result. One of the proposals to resolve it is to match Mellin moments in coordinate space and momentum space~\cite{Belitsky:2005qn,Yao:2022vtp}.  In contrast, the quark-in-gluon channel has a mixing coefficient proportional to the lightlike separation, and thus is regular when approaching the local limit and introduces no ambiguities in Fourier transform.
\label{SEC:Introduction}

On the other hand, significant progress has been made in recent years on calculating parton physics from lattice QCD (see~\cite{Cichy:2018mum,Ji:2020ect,Constantinou:2020hdm} for a review). This involves nonlocal quark and gluon operators along spacelike separations, known as quasi-light-front (quasi-LF) operators. Under an infinite Lorentz boost, the quasi-LF operators approach the LF operators defining parton physics. The flavor-singlet quark and gluon quasi-LF operators also mix in a similar way as the LF operators, but now the mismatch in mass dimensions is compensated for by the spacelike separation. A systematic study of the factorization and mixing of the quasi-LF operators has been given both in coordinate~\cite{Yao:2022vtp,Balitsky:2019krf} and in momentum~\cite{Yao:2022vtp,Ma:2022gty,Wang:2017qyg,Wang:2019tgg} space, where a similar ambiguity in Fourier transform has been observed in the gluon-in-quark channel. 

In this work, we show that the ambiguities above are due to the lack of a proper regularization prescription for the singularity that arises when the separation between gluon/quark fields in the nonlocal operators approaches zero. We then demonstrate that dimensional regularization (DR)
resolves this issue and yields consistent results in both coordinate and momentum space. This prescription is also compatible with lattice extractions of parton distributions from
nonlocal operators.

The rest of the paper is organized as follows. In Sec.~\ref{2}, we review the mixing of nonlocal gluon correlators with singlet quark correlators, in the context of evolution and factorization. In Sec.~\ref{3}, we compare the factorization formulas derived from the operator product expansion (OPE) and explicit calculations of nonlocal correlators. We also point out deficiencies in the Mellin moments matching approach. In Sec.~\ref{4}, we give a physically viable prescription for the singularity at zero field separation, with which one can obtain consistent results in coordinate and momentum space. Finally we conclude in Sec.~\ref{5}.

\section{Mixing of nonlocal gluon and quark operators} \label{2}
Throughout this paper, we focus on the mixing between gluon and quark nonlocal operators, adopting the convention in Ref.~\cite{Yao:2022vtp}
\begin{align} \label{def_op}
&{O}_{g,u}(z_1,z_2) = g^{\mu\nu}_{{\perp}} {\bf F}_{\mu\nu}\, ,&& 
{O}_{g,h}(z_1,z_2) =i \epsilon_\perp^{\mu\nu} {\bf F}_{\mu\nu}\, , \nn\\
&{O}_{q}^{\rm s}(z_1,z_2) =\frac{1}{2} \left[ O_{q}(z_1,z_2) \mp O_{q}(z_2,z_1) \right]\, ,&&
O_{q}(z_1, z_2)=\bar \psi(z_1)\Gamma [z_1, z_2] \psi(z_2)\, ,
\end{align}
where $g_{\perp}^{\mu\nu}=g^{\mu\nu}-n^{\mu}l^{\nu}-n^{\nu}l^{\mu}$, $\epsilon_{\perp}^{\mu\nu}=\epsilon^{nl\mu\nu}$, with $n$ and $l$ being lightlike vectors. ${\bf F}_{\mu\nu} = z_{12}^\rho {\rm F}_{\rho\mu}(z_1) [z_1, z_2]{\rm F}_{\nu\sigma}(z_2) z_{12}^\sigma \equiv\text F_{z_{12}\mu} (z_1) [z_1,z_2] \,\text F_{\nu z_{12}} (z_2)$ with ${\rm F}_{\mu\nu}$ being the gluon field strength tensor. $[z_1, z_2]$ denotes the Wilson line which resides in the adjoint (fundamental) representation for gluon (quark) fields, respectively. $z_{12}^\mu=z_1^\mu-z_2^\mu={\bf{z}}_1 v^\mu-{\bf{z}}_2 v^\mu={\bf{z}}_{12} v^\mu$ can be a lightcone vector (for $v^2=0$) or a spacelike vector (for $v^2=-1$). The subscript $u, h$ denotes the unpolarized and helicity gluon operator. The superscript ${\rm s}=\mp$ in $O_q^{\rm s}$ indicates the flavor-singlet quark combinations. For the unpolarized (helicity) case, one takes the minus (plus) sign in the square bracket. $\Gamma$ is the Dirac structure taken as $\gamma^+  (\gamma^+\gamma_5)$ when $v^2=0$ and $\gamma^t  (\gamma^z\gamma_5)$ when $v^2=-1$ for the unpolarized (helicity) quark operator, we have $\gamma^t=\gamma\cdot n_t,\gamma^z=\gamma\cdot n_z$, $\gamma^+=\frac{\gamma^t+\gamma^z}{\sqrt{2}},$ with $n_t^{\mu}=(1,0,0,0),n_z^{\mu}=(0,0,0,-1)$. The transversity operator is not shown here, as there is no mixing between gluon and quark transversity operators. Note that Bose symmetry indicates that the gluon operators must satisfy certain relations under $z_1\leftrightarrow z_2$. For example, $O_{g,u}(z_1,z_2)$ shall be symmetric, whereas $O_{g,h}(z_1,z_2)$ shall be anti-symmetric under the exchange $z_1\leftrightarrow z_2$.

The mixing between gluon and quark LF operators has been well studied in their scale evolution~\citep{Balitsky:1987bk}: 
\begin{equation} \label{eq:scale}
\mu\frac{d}{d\mu}O_{g,{\rm s}}(z_{1},z_{2})=\frac{1}{{\bf{z_{12}}}}\int_{0}^{1}du\int_{0}^{1}dv\,\theta(1-u-v)\mathcal{K}^{\rm s}_{gq}(u,v)O_{q}^{\rm s}(z_{12}^{u},z_{21}^{v})+\cdots\, ,
\end{equation}
where $``\cdots"$ denotes the gluon-to-gluon channel contribution and is irrelevant to the discussion in the present work. $z_{i}$ denotes the LF coordinate, and $z^u_{12}=(1-u)z_1+u z_2$. For ${\rm s}=-,+$ on the r.h.s., we have ${\rm s}=u,h$ on the l.h.s., respectively. Such a correspondence applies to the formulas below. The one-loop results have been available for a long time (see e.g.~\citep {Balitsky:1987bk}). Based on constraints from conformal symmetry, the two-loop evolution kernels are also known~\citep{Belitsky:1999hf,Braun:2019qtp}. Sandwiching the LF operators between non-forward hadron states and performing a Fourier transform, the above equation yields the evolution equation for the gluon GPD:
\begin{equation} \label{eq:LCGPD}
\mu\frac{d}{d\mu}F_{g,{\rm s}}(x,\xi,\Delta^{2})=\int_{-1}^{1}dy\,\widetilde K^{\rm s}_{gq}(x,y,\xi)F_{q}^{\rm s}(y,\xi,\Delta^{2})+\cdots.
\end{equation}
where $F_{g,\text{s}}(x,\xi,\Delta^2)$ and $F_q^{\text{s}}(x,\xi,\Delta^2)$ are defined as:
\begin{align}
F_{g,\text{s}}(x,\xi,\Delta^2)=&\int\frac{d{\bf z}_{12}}{2\pi}e^{-i{\bf z}_{12}xP^+}\big<p^{\prime}\big|O_{g,\text{s}}(z_{12},0)\big|p\big>, \nn \\
F_{q}^{\text{s}}(x,\xi,\Delta^2)=&\int\frac{d{\bf z}_{12}}{2\pi}e^{-i{\bf z}_{12}xP^+}\big<p^{\prime}\big|O_{q}^{\text{s}}(z_{12},0)\big|p\big>,
\end{align}
$x,\xi,\Delta^2$ represent the momentum fraction, skewness, and total momentum transfer squared, respectively, we have $P^+=\frac{p^{\prime,+}+p^{+}}{2}.$

Due to the $\frac{1}{{\bf z_{12}}}$ factor in Eq.\,\eqref{eq:scale} which is singular when ${\bf z_{12}}\to 0$, certain regularization procedure must be employed in order to properly Fourier transform ${\cal K}_{gq}^{\rm s}$ in eq.~\eqref{eq:scale} to its momentum space counterpart $\widetilde K_{gq}^{\rm s}$ in~\eqref{eq:LCGPD}, rendering the latter prescription-dependent.  After Fourier transforming both sides of eq.~\eqref{eq:scale} from coordinate to momentum space, we can formally write
\begin{equation} \label{a}
\widetilde K_{gq}^{\rm s}(x,y,\xi)=-i\int_{a}^{x} dx'K^{\rm s}_{gq}(x',y,\xi)\, ,
\end{equation}
with
$$K_{gq}^{\rm s}(x,y,\xi)=\int_{0}^{1}dudv\,\theta(1-u-v)\mathcal{K}^{\rm s}_{gq}(u,v)\delta((1-u-v)y+\xi(v-u)-x)\, ,$$
where $a$ is a lower integration limit~\cite{Belitsky:2005qn} that has to be fixed. A popular treatment in the literature is to separate $K^{\rm s}_{gq}(x,y,\xi)$ into different kinematic regions, and determine $a$ by imposing physical constraints~\citep{Radyushkin:1997ki,Balitsky:1997mj}. For example, in the DGLAP region, we have $y>x>\xi$,  thus one can either choose $a=\xi$  or $a=y$. However, the freedom to choose different $a$'s would yield different results, hence ambiguity still remains.

In the case of quasi-LF operators, the mixing between gluons and quarks is often formulated in the form of a factorization formula~\citep{Yao:2022vtp}:
\begin{equation} \label{origin_coor}
O_{g,{\rm s}}(z_1,z_2)=\frac{1}{\bf z_{12}}\int_{0}^{1} d\alpha d\beta\, \theta(1-\alpha-\beta) C_{gq}^{\rm s}(\alpha,\beta,\mu^2 {\bf z_{12}^2})O_q^{{\rm s},\text{l.t.}}(z_{12}^{\alpha},z_{21}^{\beta})+\cdots\, ,
\end{equation}
where $z_i$ parameterize spacelike vectors, the superscript $``\text{l.t.}"$ denotes the leading-twist projection of the operator.
$C_{gq}^{\rm s}(\alpha,\beta,\mu^2{\bf z_{12}^2})$ is the matching kernel in coordinate space. It leads to the factorization between gluon quasi-GPD and quark GPD~\citep{Yao:2022vtp,Ma:2022gty}:
\begin{equation} \label{ori_mom}
\mathbb{H}_{g,{\rm s}}(x,\xi,\Delta^2, P^z)=\int_{-1}^{1}dy\,\mathbb{C}_{gq}^{\rm s}(x,y,\xi,\frac{\mu}{P^z})H_q^{\rm s}(y,\xi,\Delta^2,\mu)+\cdots\, ,
\end{equation}
where $\mathbb{H}_{g,{\rm s}}(x,\xi,\Delta^2,P^z)$ is the gluon quasi-GPD and $H_q^{\rm s}$ is the quark LF-GPD, $P^z$ is the $z$ component of the external state's momentum, $\mathbb{C}_{gq}^{\rm s}(x,y,\xi,\frac{\mu}{P^z})$ is the matching kernel in the momentum space which can be obtained from $C_{gq}^{\rm s}$ as follows:
\begin{equation} \label{eq:rela}
\mathbb{C}_{gq}^{\rm s}(x,y,\xi,\frac{\mu}{P^z})=\int\frac{d{\bf z_{12}}}{2\pi}\frac{1}{\bf z_{12}}\int_{\alpha\beta}C_{gq}^{\rm s}(\alpha,\beta,\mu^2{\bf z_{12}^2})\, ,
\end{equation}
where we have used the shorthand notation $$\int_{\alpha\beta}\equiv\int_{0}^{1} d\alpha d\beta\, \theta(1-\alpha-\beta)e^{i{\bf z_{12}}P^z\big(x+\xi(-\alpha+\beta)+(1-\alpha-\beta)y\big)}\, .$$ 
Note that Eq.\,\eqref{eq:rela} involves the Fourier transform of both $\frac{1}{\bf z_{12}}$ and $\frac{\text{ln}{\bf z_{12}^2}}{{\bf z_{12}}}$, which, without specifying a proper prescription for the singularity at ${\bf z_{12}}=0$, will lead to an ambiguous/singular result. For example, we have for the Fourier transform $I(x)\equiv\int\frac{dz}{2\pi}e^{izP^zx}\frac{1}{z}=i\frac{\epsilon(x)}{2}+C,$ where the indefinite constant $C$ is obtained by first differentiating $I(x)$ with respect to $x$ and then integrating back.

From the discussion above, the ambiguity in the mixing channel contribution in momentum space shares a similar origin for both quasi-LF operators and LF operators: the absence of a well-defined prescription for handling the singularity at ${\bf z_{12}=0}$ in Fourier transform. 
In the following, we will focus on the quasi-LF operators, since once a consistent regularization prescription for the singularity at ${\bf z_{12}=0}$ is established, the ambiguity in Eq.\,\eqref{a} can be eliminated in the same way, due to the quasi-LF factorization requirement in the momentum space: the $\mu$-dependence of the LF-GPDs must match the scale-dependence of $P^z$ on quasi-LF side.

\section{The role of \texorpdfstring{$\frac{1}{\bf z_{12}}$}{z12} in the Fourier transform} \label{3}
The factorization for the quasi-LF correlator
is derived at a nonzero separation of the gluon and quark fields.
However, as mentioned in the previous section, when Fourier transformed to momentum space, we need a well-defined prescription to regularize the pole at ${\bf z_{12}=0}$. To illustrate the problem, we compare in Sec.~\ref{SEC:operator} the difference between the traditional operator product expansion (OPE) approach and perturbative factorization in coordinate space. For simplicity, we use the forward gluon correlator as an example. In Sec.~\ref{SEC:MMmatching}, we point out the deficiency of a certain approach: matching Mellin moments of the matching kernels in coordinate and momentum space.

\subsection{Traditional OPE and factorization in coordinate space}
\label{SEC:operator}
Let us review briefly the traditional OPE for the mixing channel of the gluon quasi-LF correlator (forward case)\cite{Wang:2019tgg}: 
\begin{equation} \label{eq:OPEref}
\tilde{h}_{g,{\rm s}}({\bf z_{12}^{2}},\lambda)=(P^{z})^{2}\sum_{n=2}^{\infty}\frac{(-i\lambda)^{n-2}}{(n-2)!}C_{n-2}^{gq,{\rm s}}(\mu^{2}{\bf z_{12}^{2}})a_{n-1}^{q,{\rm s}}(\mu)+\cdots,
\end{equation}
with $\lambda={\bf z_{12}}P^z$ denoting the quasi-LF distance. The correlator $\tilde{h}_{g,{\rm s}}({\bf z_{12}^2},\lambda)$ is defined as:
\begin{equation}
\tilde{h}_{g,{\rm s}}({\bf z_{12}^2},\lambda)=\big<P\big|O_{g,{\rm s}}(z_1,z_2)\big|P\big>,
\end{equation}
with $O_{g,{\rm s}}(z_1,z_2)$ given in Eq.\,\eqref{def_op}. The polarization vector $\vec S$ of the external nucleon state is omitted for brevity.
We use the following definitions of quark singlet distribution and moment $a_{n}^{q,{\rm s}}$:
\begin{align}
f_q^{\rm s}(y)=&\int\frac{d{\bf z}_{12}}{2\pi}e^{-i{\bf z}_{12}P^+y}\big<P\big|O_q^{\rm s}(z_{12},0)\big|P\big>, 
\qquad
a_{n}^{q,{\rm s}}=\int_{-1}^{1} dyy^nf_q^{\rm s}(y).
\end{align}
Making a Fourier transform on both sides of Eq.\,\eqref{eq:OPEref}, we obtain the momentum space factorization formula for the gluon quasi-PDF: 
\begin{equation} \label{fact_PDF}
x\tilde{f}_{g,{\rm s}}(x,P_z)=\int_{-1}^{1}dy\,\mathbb{C}_{gq}^{\text{OPE},{\rm s}}(x,y,\frac{\mu}{P^{z}})f_{q}^{\rm s}(y)+\cdots,
\end{equation}
where
\begin{equation}\label{eq:xf-tilde}
x\tilde{f}_{g,{\rm s}}(x,P_z)=\int\frac{d{\bf z_{12}}}{2\pi P^z}e^{ix{\bf z_{12}}P^z}\tilde{h}_{g,{\rm s}}({\bf z_{12}^2},\lambda),
\end{equation}
\begin{equation} \label{re_OPE}
\mathbb{C}_{gq}^{\text{OPE},{\rm s}}(x,y,\frac{\mu}{P^{z}})=P^z\int\frac{d{\bf z_{12}}}{2\pi}e^{ix{\bf z_{12}}P^z}\sum_{n=2}^{\infty}\frac{(-i{\bf z_{12}}P^z)^{n-2}}{(n-2)!}C_{n-2}^{gq,{\rm s}}(\mu^{2}{\bf z_{12}^{2}})y^{n-1},
\end{equation}
$\mathbb{C}_{gq}^{\text{OPE},{\rm s}}$ doesn't include the Fourier transform of $\frac{1}{{\bf z_{12}}}$. 

On the other hand, performing a local expansion for the factorization 
formula matching directly between the nonlocal quasi-LF and LF correlators, we have,
\begin{equation} \label{eq:coor_direct}
\tilde{h}_{g,{\rm s}}({\bf z_{12}^{2}},\lambda)
=\frac{\sqrt{2}P^{z}}{{\bf z_{12}}}\sum_{n=2}^{\infty}\frac{(-i\lambda)^{n-2}}{(n-2)!}\int_{0}^{1}d\alpha\,\alpha^{n-2}C_{gq}^{\text{coord},{\rm s}}(\alpha,\mu^{2}{\bf z_{12}^{2}})a_{n-2}^{q,\rm s}(\mu)+\cdots.
\end{equation}
 In Eq.\,\eqref{eq:coor_direct}, we have used the superscript $\text{``coord''}$ to distinguish the matching kernel from that obtained from OPE:
 $C_{gq}^{\text{OPE},\rm s}$.
 For the unpolarized case, the term $\propto\frac{1}{{\bf z_{12}}}$ in Eq.\,\eqref{eq:coor_direct} vanishes because the first moment of the singlet quark distribution is zero (i.e., $a_0^{q,-}=0$),
 Eq.\,\eqref{eq:OPEref} and Eq.\,\eqref{eq:coor_direct} are therefore equivalent. For the polarized case, the equivalence condition becomes $\int_{0}^{1}d\alpha\, C_{gq}^{\text{coord},\rm s}(\alpha,\mu^2{\bf z_{12}^2})=0$ due to $a_0^{q,+} \neq 0$. 
 For the nonforward case, the equivalence condition reads $\int_{\alpha\beta}C_{gq}^{\text{coord},\rm s}(\alpha,\beta,\mu^2{\bf z_{12}^2})=0$. 
 Only one-loop matching kernel $C_{gq}^{\text{coord},\rm s}$ is available in the literature, 
 it can be separated into the evolution part ($C_{gq}^{\text{evol},\rm s}$) and the 
 scale-independent part ($C_{gq}^{\text{si},\rm s}$):
\begin{equation} \label{structure}
C_{gq}^{\text{coord},\rm s,(1)}(\alpha,\mu^2{\bf z_{12}^2})=C_{gq}^{\text{si},\rm s}(\alpha)+C_{gq}^{\text{evol},\rm s}(\alpha)\text{ln} \left(\frac{{\bf z_{12}^2}\mu^2}{4e^{-2\gamma_E}}\right).
\end{equation}
From explicit calculations~\citep{Yao:2022vtp}, we find that $C_{gq,h}^{\text{evol},\rm s}$ has a vanishing first moment at one-loop:
\begin{equation} \label{first}
\int_{0}^1d\alpha\, C_{gq,h}^{\text{evol},\rm s}(\alpha)=0,
\end{equation}
however, no symmetry ensures that this holds at higher orders in perturbation theory.
Moreover, the first moment of $C_{gq,h}^{\text{si},\rm s}(\alpha)$ is nonzero. 


\subsection{Matching of Mellin moments: deficiency}
\label{SEC:MMmatching}
A possible proposal for removing the ambiguities in the Fourier transform Eq.\,\eqref{eq:rela} is to match Mellin moments of the matching coefficient in momentum space with those in coordinate space~\cite{Belitsky:2005qn,Yao:2022vtp}. Consider the following matching formula~\citep{Yao:2022vtp}:
\begin{equation} \label{matching_master}
\lim\limits_{\xi\to0}\int_0^1dx \, x^{j}\mathscr{C}_{gq}^{\rm s}\left(2\frac{x}{\xi}-1,\frac{2}{\xi}-1\right)=\int_{0}^{1}d\alpha\, C_{gq}^{\rm s}(\alpha,\mu^{2}{\bf z_{12}^{2}})\frac{-\bar{\alpha}^{j+1}}{j+1},\quad j\geq0\, , 
\end{equation}
where $\bar\alpha = 1-\alpha$. 
Eq.\,\eqref{matching_master} is based on the relation between the evolution kernel in DGLAP and ERBL region (see Eq. (4.93) in \citep{Belitsky:2005qn}). 
Since $C_{gq}^{\rm s}(\alpha,\mu^{2}{\bf z_{12}^{2}})$ on the r.h.s. is known, the task then becomes finding $\mathscr{C}_{gq}^{\rm s}\left(2\frac{x}{\xi}-1,\frac{2}{\xi}-1\right)$ on the l.h.s. such that the above equation is satified.
Here $\mathscr{C}_{gq}^{\rm s}$ can be either the evolution kernel $\widetilde K^{\rm s}_{gq}$ (Eq.~\eqref{eq:LCGPD}, for GPD)~\citep{Belitsky:2005qn} or the matching coefficient function (for pseudo-GPD)~\citep{Yao:2022vtp}. 

However, such a method is not sufficient to eliminate all the ambiguities in the Fourier transform, due to the following two facts:
\begin{itemize}
\item The evolution kernel of gluon PDF (or matching kernel of pseudo gluon-PDF) can also be obtained through a Fourier transform of the kernel in coordinate space, its moments can be explicitly calculated:
\begin{equation} \label{nonfor-coor}
\int_0^1 dx x^j \mathscr{C}_{gq}^{\rm s}(x,y,\mu^2 \mathbf{z_{12}^2})=\int_{0}^{1}d\alpha\, C_{gq}^{\rm s}(\alpha,\mu^{2}{\bf z_{12}^{2}})\Big(\frac{-(\bar{\alpha}y)^{j+1}+1}{j+1}+\frac{C(\bar{\alpha}y)}{j+1}\Big),\quad j\geq0,
\end{equation}
where we have replaced the inverse correlation length with an indefinite integral, $C(\bar{\alpha}y)$ on the r.h.s. is the integration constant. Comparing Eq.~\eqref{nonfor-coor} with Eq.~\eqref{matching_master}, we find that Eq.~\eqref{matching_master} implicitly assumes $C(\bar{\alpha}y)=-1$. 
However, its physical relevance is yet to be justified. 
\item The solution $\mathscr{C}_{gq}^{\rm s}$ satisfying Eq.\,\eqref{matching_master} is not unique, since one only fixes the DGLAP region in Eq.\,\eqref{matching_master}. Under the requirement of charge conjugation symmetry, 
one can easily construct different functions of $\mathscr{C}_{gq}^{\rm s}(x,y)$ that are not equivalent at the convolution level.
\end{itemize}


\section{Regularization prescription for the \texorpdfstring{$\frac{1}{\bf z_{12}}$}{1z12} pole}
\label{4}
The matching kernels in coordinate space contain contributions singular as ${\bf z_{12}}\rightarrow0$, preventing us from getting a smooth local limit of the factorization. Correspondingly, the first Mellin moments of quasi-PDFs and quasi-GPDs are divergent. These divergences originate from the non-commutativity of the two limits: $P^z\rightarrow\infty$ (in momentum space)/${\bf z_{12}^2}\rightarrow0$ (in coordinate space) and $\Lambda_{\rm UV}\rightarrow\infty$. In this section, we show that dimensional regularization (DR) with $d=4-2\epsilon$ can properly treat these divergences, yielding a physically viable prescription for the singularity in practical calculations. We also discuss the principal-value prescription for comparison.

\subsection{Recovery of a smooth local limit}

For illustration purposes, we first consider the factorization of forward kinematics in coordinate space at one-loop, in the $\overline{\text{MS}}$ scheme. For the unpolarized case, taking the local limit of the factorization formula in Eq.\,\eqref{origin_coor}, we get:
\begin{align}\label{bare_unpol}
\lim_{{\bf z_{12}}\rightarrow0}\big<P\big|O_{g,u}(z_{12},0)\big|P\big>\!=\!-2ia_{s}C_{F}\lim_{{\bf z_{12}}\rightarrow0} &\int_{0}^{1}d\alpha\left[6\alpha-4+\delta(\alpha)-\text{ln}\left(\frac{{\bf z_{12}^2}\mu^2}{4e^{-2\gamma_E}}\right)(2\alpha+\delta(\alpha))\right]\notag\\
&\times(1-\alpha)\big<P\big|O_{q}^{-,1}(0,0)\big|P\big>,
\end{align}
where $a_s=\alpha_s/(4\pi)$. 
In the r.h.s. of Eq.\,\eqref{bare_unpol}, we have made a tree-level expansion: $$\langle P | O_{q}^{-}(z_{12},0) | P\rangle={\bf z_{12}}\langle P | O_{q}^{-,1}(0,0) | P\rangle+\mathcal{O}({\bf z_{12}^3}),$$ which is sufficient for our purpose at $\mathcal{O}(\alpha_s)$, the fact that $\langle P | O_{q}^{-}(z_{12},0)| P\rangle$ is an \textbf{odd} function in ${\bf z_{12}}$~\citep{Balitsky:2019krf} is used. 
Here, $O_{q}^{-,1}(0,0)$ is the unpolarized local quark operator with one covariant derivative whose explicit expression doesn't concern us.
The r.h.s. of Eq.\,\eqref{bare_unpol} is at $\mathcal{O}({\bf z_{12}^0})$, 
there still remains a term $\propto\text{ln}(\mu^2\bf z_{12}^2)$, which is singular as ${\bf z_{12}}\rightarrow0$. In the spirit of DR, we can make the replacement in $\overline{\text{MS}}$ scheme$$\text{ln}\frac{{\bf z_{12}^2}\mu^2}{4e^{-2\gamma_E}}\rightarrow\frac{(\frac{{\bf z_{12}^2}\mu^2}{4e^{-2\gamma_E}})^{\epsilon}-1}{\epsilon}$$ 
in Eq.~\eqref{bare_unpol}, where we take $\epsilon>0$ to regulate the UV divergence as $\bf z_{12}\rightarrow 0$. Evaluating the integration over $\alpha$ on the r.h.s., Eq.~\eqref{bare_unpol} becomes:

\begin{equation} \label{re_unpol}
\big<P\big|O_{g,u}(0,0)\big|P\big>=-\frac{8ia_{s}C_{F}}{3\epsilon_{\text{UV}}}\big<P\big|O_{q}^{-,1}(0,0)\big|P\big>\, .
\end{equation}
Physically, Eq.\,\eqref{re_unpol} is just the mixing of gluon average momentum with quark average momentum. 
The coefficient of the UV pole in Eq.\,\eqref{re_unpol} agrees with that calculated using the local matrix element $\big<P\big|O_{g,u}(0,0)\big|P\big>$.

For the polarized case, $\langle P | O_{q}^{+}(z_{12},0) | P\rangle$ is \textbf{even} in ${\bf z_{12}}$ 
and therefore the r.h.s. of Eq.\,\eqref{origin_coor} has a $\frac{1}{{\bf z_{12}}}$ divergence as ${\bf z_{12}}\rightarrow0$.
We can again employ DR to regulate this divergence, the local limit of Eq.\,\eqref{origin_coor} becomes:
\begin{equation} \label{bare_pol}
    \lim_{{\bf z_{12}}\rightarrow0}\big<P\big|O_{g,h}(z_{12},0)\big|P\big>= \lim_{{\bf z_{12}}\rightarrow0}\frac{6ia_sC_F({\bf z_{12}^2})^{\epsilon}}{{\bf z_{12}}}g_A,
\end{equation}
The r.h.s. of Eq.\,\eqref{bare_pol} is free from the logarithmic term since we have used $\int_{0}^1d\alpha \;C_{gq,h}^{\text{evol}}(\alpha)=0$ at ${\cal O}(\alpha_s)$, prior to taking the local limit. For $\epsilon>\frac{1}{2}$, the r.h.s. is zero and consistent with the l.h.s., using the fact that $O_{g,h}(z_{12},0)$ is \textbf{odd} in ${\bf z_{12}}$. The region of convergence on the r.h.s. of Eq.~\eqref{bare_pol} can be extended beyond $\epsilon>1/2$ by analytic continuation of $\epsilon$.

We then consider the factorization of gluon quasi-PDF in momentum space.
We use the following factorization in coordinate space in $d=4-2\epsilon$ dimensions:
\begin{equation} \label{D_coor}
O_{g,{\rm s}}(z_{12},0)=\frac{(\bf z_{12}^2)^{\epsilon}}{\bf z_{12}}\int_{0}^{1} d\alpha C_{gq}^{\rm s}(\alpha,\mu^2, \epsilon)O_q^{{\rm s},\text{l.t.}}((1-\alpha)z_{12},0)+\cdots, 
\end{equation}
the expression of $C_{gq}^{\rm s}(\alpha,\mu^2, \epsilon)$ can be read from Eq.(3.11) in \cite{Yao:2022vtp}, the matching kernel in momentum space in $4-2\epsilon$ dimensions can be obtained as follows:
\begin{equation} \label{eq:reladim}
\mathbb{C}_{gq}^{\rm s}(x,y,\frac{\mu}{P^z},\epsilon)=\int\frac{d{\bf z_{12}}}{2\pi}\frac{({\bf z_{12}^2})^{\epsilon}}{{\bf z_{12}}}
\int_{0}^{1} d\alpha \,e^{i{\bf z_{12}}P^z(x+(1-\alpha)y)}\,C_{gq}^{\rm s}(\alpha,\mu^2,\epsilon),
\end{equation}
$\mathbb{C}_{gq}^{\rm s}(x,y,\frac{\mu}{P^z},\epsilon)$ has support $x\in (-\infty,\infty)$,
it has the following structure:
\begin{align} \label{dim_mom}
\mathbb{C}_{gq}^{\rm s}(x,y,\frac{\mu}{P^z},\epsilon)=&a(x,y)\frac1\epsilon\left[1-\left(\frac{4(P^z)^2}{\mu^2}\right)^{-\epsilon}\right] \nonumber\\
&+d(x,y,\epsilon)\big|x-y\big|^{1-2\epsilon}+e(x,y,\epsilon)\big|x\big|^{1-2\epsilon}+f(x,y,\epsilon)\big|x-y\big|^{-2\epsilon},
\end{align}
where $d,e,f$ are polynomials in $x,y,\epsilon$. 
Taking the infinite momentum limit,
the $P^z$-dependence in Eq.\,\eqref{dim_mom} disappears for positive $\epsilon$.
We can then calculate the second moment $\int_{-\infty}^\infty dx\,x\,\tilde f(x,\infty)$ of $\tilde{f}_{g,\rm s}(x,\infty)$ from Eq.~\eqref{fact_PDF} in $4-2\epsilon$ dimensions, the results read:
\begin{align} \label{moment_mom}
\big<P\big|O_{g,u}(0,0)\big|P\big>&=-\frac{8ia_sC_F}{3\epsilon_{\text{UV}}}\big<P\big|O_{q}^{-,1}(0,0)\big|P\big>,\quad\text{for the unpolarized case}, \nonumber \\
0&=0,\quad\text{for the polarized case}.
\end{align}
Eq.\,\eqref{moment_mom} is consistent with the local limit (OPE) of spacelike correlators, with DR adopted. 
Eq.~\eqref{moment_mom} is obtained by evaluating the second moment of $\mathbb{C}_{gq}^{\rm s}(x,y,\frac{\mu}{P^z},\epsilon)$:
the integration of the second line in Eq.\,\eqref{dim_mom} can be written as a sum of 
scaleless integrals after proper shifts of the integration variable $x$, which vanish in DR. Thus, the moment of the finite term in Eq.\,\eqref{dim_mom} is zero.
Note that in 4 dimensions, the moment of $\mathbb{C}_{gq}^{\rm s}(x,y,\frac{\mu}{P^z},0)$ in Eq.~\eqref{dim_mom} is divergent~\citep{Izubuchi:2018srq,Ma:2022gty}. 

The above discussions focus on the forward correlators and quasi-PDFs. For the nonforward correlators and quasi-GPDs, employing DR, one can correctly reproduce the mixing patterns of local nonforward matrix elements, consistent with the results via direct calculations. It is worth noting that certain polarized gluon quasi-PDFs have vanishing first moments in the $\overline{\text{MS}}$ scheme as well to all orders, due to their relation with the matrix element of topological current~\citep{Pang:2024sdl,Hatta:2013gta}.

In lattice calculations of nonlocal correlators, the results are smooth when approaching the local limit (${\bf z_1}\to {\bf z_2}$). Since here one implicitly includes contributions from all orders in $\alpha_s$, the singular terms such as $\ln {\bf z_{12}^2}$ in fixed-order perturbation theory is effectively resummed, giving a structure expected to be polynomial in ${\bf z_{12}}$~\citep{Su:2022fiu}. In this sense, DR is also consistent with resummation through solving the renormalization group equation.

\subsection{Comparison of two methods for calculating \texorpdfstring{$\mathbb{C}_{gq}^{\rm s}$}{Cgq}}

In the previous subsection, we have shown that DR yields a consistent regularization prescription for the $1/{\bf z_{12}}$ pole. We denote the expansion of $\mathbb{C}_{gq}^{\rm s}(x,y,\frac{\mu}{P^z},\epsilon)$ (in Eq.\,\eqref{eq:reladim}, for GPD it is $\mathbb{C}_{gq}^{\rm s}(x,y,\xi,\frac{\mu}{P^z},\epsilon)$) at $\epsilon=0$ as $\mathbb{C}_{gq}^{\text{FT,DR},\rm s}$.
Their expressions read: 
\begin{align}
\mathbb{C}_{gq}^{\text{FT,DR},\rm u}(x,y,\frac{P_z}{\mu})=&-4a_sC_F\Big\{\text{sgn}(x-y)\frac{(x-y)^2+y^2}{2y^2}\text{ln}\frac{4P_z^2(x-y)^2}{\mu^2}-\text{sgn}(x)\frac{x(x-2y)}{2y^2}\text{ln}\frac{4P_z^2x^2}{\mu^2}+\text{sgn}(x-y)\frac{4x-3y}{2y} \nn \\
&-\text{sgn}(x)\frac{x}{y}\Big\}, \nn\\
\mathbb{C}_{gq}^{\text{FT,DR},\rm h}(x,y,\frac{P_z}{\mu})=&-4a_sC_F\Big\{\frac{x(x-2y)}{2y^2}\Big(\text{sgn}(x)\text{ln}\frac{4P_z^2x^2}{\mu^2}-\text{sgn}(x-y)\text{ln}\frac{4P_z^2(x-y)^2}{\mu^2}\Big)+\text{sgn}(x)\frac{x(y-2x)}{y^2} \nn\\
&+\text{sgn}(x-y)\frac{2x(x-y)}{y^2}\Big\},
\end{align}
for matching kernels of quasi-PDFs, and
\begin{align}
\mathbb{C}_{gq}^{\text{FT,DR},\rm u}(x,y,\xi,\frac{P_z}{\mu})=&-2a_sC_F\Big\{\Big(\text{sgn}(x+\xi)\frac{(x+\xi)^2}{2\xi(y+\xi)}\text{ln}\frac{4(P_z)^2(x+\xi)^2}{\mu^2}-\text{sgn}(x+\xi)\frac{2(x+\xi)}{y+\xi} \nn\\
&+\text{sgn}(x-y)\frac{x^2-2xy+2y^2-\xi^2}{2(y^2-\xi^2)}\text{ln}\frac{4(P_z)^2(x-y)^2}{\mu^2}-\text{sgn}(x-y)\frac{\xi^2-4xy+3y^2}{2(y^2-\xi^2)}\Big)\nn\\
&+(\xi\rightarrow-\xi)\Big\},\nn\\
\mathbb{C}_{gq}^{\text{FT,DR},\rm h}(x,y,\xi,\frac{P_z}{\mu})=&-2a_sC_F\Big\{\Big(-\text{sgn}(x+\xi)\frac{(x+\xi)^2}{2\xi(y+\xi)}\text{ln}\frac{4(P_z)^2(x+\xi)^2}{\mu^2}+\text{sgn}(x+\xi)\frac{2x(x+\xi)}{\xi(y+\xi)}\nn\\
&-\text{sgn}(x-y)\frac{\xi^2+x^2-2xy}{2(y^2-\xi^2)}\text{ln}\frac{4(P_z)^2(x-y)^2}{\mu^2}+\text{sgn}(x-y)\frac{2x(x-y)}{y^2-\xi^2}\Big) \nn\\
&+(\xi\rightarrow-\xi)\Big\},
\end{align}
for matching kernels of quasi-GPDs.

Another method to calculate $\mathbb{C}_{gq}^{\rm s}$ is to first Fourier transform the amplitude to momentum space, which implicitly includes the contribution at ${\bf z_{12}=0}$, then work out the loop integral in momentum space~\citep{Ma:2022gty,Wang:2017qyg}, this results are denoted as $\mathbb{C}_{gq}^{\text{mom},\rm s}$.

In Table \ref{Tab:comparison_directFT}, we compare the two sets of matching kernels at one-loop: $\mathbb{C}_{gq}^{\text{FT,DR},\rm s}$ and $\mathbb{C}_{gq}^{\text{mom},\rm s}$ (in~\citep{Ma:2022gty}), of quasi-PDFs and quasi-GPDs. They are not the same superficially, but the differences don't contribute after convoluting with the LF-distributions due to the symmetry properties of quark singlet distributions (in Eq.\,\eqref{ori_mom} and Eq.\,\eqref{fact_PDF}).

Note that in ~\citep{Ma:2022gty}, extra minus signs should be added to the gluon-in-quark matching kernels for quasi-PDFs (Eq.(4.13) and Eq.(4.15) in ~\citep{Ma:2022gty}), after which the results are consistent with DGLAP evolutions. 

The gluon-in-quark mixing channel receives contributions from two diagrams, which are crossed to each other. In coordinate space, integration-by-parts (IBPs) over the Feynman parameters $\alpha,\beta$ is used to eliminate certain contributions 
causing subtleties for later coordinate $\mapsto$ momentum space Fourier transformation (see Eq.\,(3.10)---(3.12) in~\citep{Yao:2022vtp}). On the other hand, for
all diagrams, making a Fourier transform of the direct diagrammatic coordinate space result \textit{before} the IBPs (Eq.\,(3.10) in~\citep{Yao:2022vtp}), we can arrive at an expression exactly the same as that directly calculated in momentum space. It is interesting to note that besides the gluon-in-quark kernel, the momentum space parity-even quark-in-gluon evolution kernel (for quark GPD) and gluon-in-gluon matching kernel (for quasi gluon-PDF) are not unique in the literature either ~\citep{Belitsky:1998vj,Diehl:2003ny,Ji:1996nm,Radyushkin:1997ki,Balitsky:2019krf,Yao:2022vtp,Ma:2022gty}. The differences can be attributed to the adoption of the following two computational methods:
\begin{itemize}
    \item Calculate the kernel directly in momentum space,
    \item First derive the evolution in coordinate space with IBP applied, then Fourier transform to momentum space.
\end{itemize} 
We have explicitly checked that the differences in ~\citep{Belitsky:1998vj,Diehl:2003ny,Ji:1996nm,Radyushkin:1997ki,Balitsky:2019krf,Yao:2022vtp,Ma:2022gty} are irrelevant as their contribution
vanishes at the level of momentum space factorization theorem. Therefore, they have no impact on physical results. Such
``superficial” differences are likely attributable to the use of IBPs in coordinate space. We leave for a future study whether one can obtain exactly the same kernel (or
matching coefficient) as that directly calculated in momentum space if one Fourier transforms the evolution kernel
(or matching coefficient) before applying IBP in the coordinate space.

\begin{table}[htbp]
  \centering
  \renewcommand\arraystretch{1.5}
  \begin{tabular}{ccc}
\hline
~~~ & $\text{evolution}$ & $\text{finite part}$  \\
\hline
\specialrule{0em}{2pt}{2pt}
unpolarized & 
$\checkmark$ & $\checkmark$ \\
\specialrule{0em}{2pt}{2pt}
polarized & $\checkmark$ & $\checkmark$ \\
\specialrule{0em}{2pt}{2pt}
\hline
\end{tabular}
\caption{The comparison between $\mathbb{C}_{gq}^{\text{mom},\rm s}$ and $\mathbb{C}_{gq}^{\text{FT,DR},\rm s}$, the ``evolution" in the head row means the coefficients of $\text{ln}\frac{4(P^z)^2}{\mu^2}$. The comparison is performed at the convolution level.
The results in the table hold for both quasi-PDFs and quasi-GPDs.}
 \label{Tab:comparison_directFT}
\end{table}
Alternatively, one may think of using the principal-value prescription to regulate the poles at ${\bf z_{12}}=0$ in Eq.\,\eqref{bare_unpol} and Eq.\,\eqref{bare_pol} (in ~\citep{Blumlein:1999sc}, the principal-value prescription was also used to calculate the evolution of gluon GPD to quark GPD).
It is worth pointing out that at one-loop, the $\mathbb{C}_{gq}^{\rm s}$ calculated using principal-value prescription: 
\begin{equation} \label{eq:relaPV}
\mathbb{C}_{gq}^{\text{FT,PV},\rm s}(x,y,\xi,\frac{\mu}{P^z})=\int\frac{d{\bf z_{12}}}{2\pi}\text{P}\Big[\frac{1}{{\bf z_{12}}}\int_{\alpha\beta}C_{gq}^{\rm s}(\alpha,\beta,\mu^2{\bf z_{12}^2})\Big],
\end{equation}
agrees with $\mathbb{C}_{gq}^{\text{FT,DR},\rm s},$ for all types of quasi distributions. The operation $``\text{P}\left[f({\bf z_{12}})\right]"$ means subtracting the point at ${\bf z_{12}}=0$ in $f({\bf z_{12}})$.
But unlike DR, in some cases, 
 convoluting the LF operator with the coefficient function regularized using the principal-value prescription gives inconsistent results in the local limit.  Employing the factorization theorem for the polarized forward correlator, and using principal-value prescription to obtain the matching coefficient function, one gets a consistent result 
in the local limit.
However, the factorization theorem for the polarized nonforward correlator implies the following local limit:
\begin{align} \label{nonforward-pol}
0=&\int_0^1d\alpha\int_0^{1-\beta}C_{gq}^{+}(\alpha,\beta,\mu^2{\bf z_{12}^2})(1-\alpha+\beta)\big<P'\big|O_{q}^{+,1}(0,0)\big|P\big> \nonumber \\
=&\;6ia_sC_F\big<P'\big|O_{q}^{+,1}(0,0)\big|P\big>,
\end{align}
if principal-value prescription is used. Eq.\,\eqref{nonforward-pol} 
is evidently inconsistent as 
$$\langle P'\big|O_{q}^{+,1}(0,0)\big|P\rangle \neq 0\, .$$
Moreover, in momentum space, we haven't found a direct implementation of principal-value prescription, such that gluon quasi-PDFs and GPDs have finite first moments.

In a recent work~\citep{Yao:2022vtp}, the mixing channel contributions to the momentum space gluon quasi-GPDs and PDFs are conveniently determined by matching the Mellin moments in coordinate and momentum space. The results turn out to be equivalent to those obtained in the present work, except for the logarithm-independent part in the gluon-in-quark channel (dictated by the matching coefficient ${\mathbb C}_{gq}$) for the gluon helicity quasi-GPD and quasi-PDF, which requires a minor revision. 
For calculations of the gluon-in-quark channel directly in momentum space, one can refer to~\citep{Ma:2022gty} (for quasi-GPDs and quasi-PDFs) and~\citep{Wang:2019tgg} (for quasi-PDFs), where the results are consistent with the conclusions in this work.

\section{Conclusion} \label{5}
In this work, we present a systematic investigation of the mixing between nonlocal gluon operators and flavor-singlet quark operators for both lightlike and spacelike separations. Our analysis reveals that dimensional regularization provides a simple, unifying and physically viable prescription for regularizing the pole at zero field separation, which also reproduces the expected mixing pattern of local operators. Through explicit calculations, we demonstrate that the coordinate space factorization at one-loop, when treated with DR and Fourier transformed to momentum space, yields results consistent with those obtained from direct calculations in momentum space\footnote{Namely, it is necessary to Fourier transform the $ d$-dimensional coordinate space matching coefficient function and  perform the $\epsilon$ expansion for the obtained momentum space result at the very last step.}. Our work resolves a long-standing challenge in relating the coordinate and momentum space results for nonlocal operator mixing, and is expected to have a significant impact on lattice calculations of gluon PDFs (GPDs) and singlet quark PDFs (GPDs) within the LaMET framework. 
\label{SEC:conclusion}

\vspace{2em}

\acknowledgments 
We thank Andrei V. Belitsky, Vladimir M. Braun for valuable discussions on 
moments matching and the singular behavior of the gluon-in-quark channel as ${\bf z_{12}}\rightarrow0$. 
We also thank Guangpeng Zhang for helpful discussions and comparing evolution kernels obtained using different methods.
This work is supported in part by National Natural Science Foundation of China under grants No. 12375080, 11975051, No. 12061131006, the Ministry of Science and Technology of China under Grant No. 2024YFA1611004, and by CUHK-Shenzhen under grant No. UDF01002851 and UDF01003869. FY is partially supported by the U.S. Department of Energy, Office of Science, Office of Nuclear Physics through Contract No. DE-SC0012704, and within the framework of Scientific Discovery through Advanced Computing (SciDAC) award Fundamental Nuclear Physics at the Exascale and Beyond.



\bibliography{apssamp}

\end{document}